# On Degenerate Metrics and Electromagnetism

T.P. Searight

*Toronto-Dominion Bank, Triton Court, 14/18 Finsbury Square,*
*London EC2A 1DB, United Kingdom*
*e-mail: trevor.searight@tdsecurities.com*

**Abstract**

A theory of degenerate metrics is developed and applied to the problem of unifying gravitation with electromagnetism. The approach is similar to the Kaluza-Klein approach with a fifth dimension, however no ad hoc conditions are needed to explain why the extra dimension is not directly observable under everyday conditions. Maxwell's theory is recovered with differences only at very small length scales, and a new formula is found for the Coulomb potential that is regular everywhere.

## 1  Introduction

It is an old question in physics: Is there an underlying reason why Newton's law of gravitation and Coulomb's law both follow the inverse square rule? The Kaluza-Klein theory [1, 2] where electromagnetism is described by curvature in an extra spacelike dimension is the best-known solution to this problem, however it has an obvious flaw: we do not experience a fifth dimension as part of our everyday lives. An extra condition is therefore introduced to the theory to explain away this difficulty: the fifth dimension is supposed to be "curled up small" so that it is not noticeable except at small length scales. There is also a restriction on the set of transformations under which the theory is invariant. These peculiarities suggest that Kaluza-Klein may not be the correct theory (general relativity has no such additional properties). Instead I will use a degenerate metric to argue that there could be more than the usual four dimensions to the universe.

There have already been various studies of degenerate metrics. Much of the activity in this field uses the Ashtekar formulation of general relativity [3] which is polynominal in the canonical variables and so permits a degenerate extension. Several authors have investigated a degenerate metric which is the boundary between two non-degenerate regions (see [4] and the references therein). The motion of particles and strings in a degenerate metric space has also been investigated [5].

In my approach, the metric is degenerate at all points, not just in some region or at some boundary. Such a metric describes a new type of dimension which is neither spacelike nor timelike, which I shall call *lightlike* or *null*. Because there are similarities between the degenerate extension and the Kaluza-Klein theory I will work in five dimensions, with a single lightlike dimension plus the usual four space-time dimensions. However it should be noted that the work in this paper can also be considered solely in the context of a degenerate extension to relativity without reference to electromagnetism or unified field theory. Throughout I will demand invariance under all coordinate transformations, on the basis that general relativity is invariant under all coordinate transformations.

I begin in section 2 by defining the covariant and contravariant metrics as far as is possible using algebra when one or other is degenerate. In section 3 tensor calculus for degenerate metrics is developed along the lines of general relativity with Christoffel symbols and Riemann and Ricci tensors. All the scalars which can be used to build a Lagrangian are also found. Two possible equations of motion for a point particle are discussed in section 4, and Maxwell's theory appears with one of them. In section 5 the field equations are developed and again Maxwell's equations appear, however at first sight there is a problem: charges do not act directly as a source for electromagnetism (they are only acted on by it) and therefore Coulomb's law cannot immediately be retrieved. This problem is resolved in section 6 using the non-linearity of the field equations: electromagnetism can act as its own source. A spherically symmetric solution is found and a new formula for the Coulomb potential that is regular everywhere. There are some closing comments in section 7.

## 2   Defining the Metric

So how does a degenerate metric allow a new type of dimension? In general relativity a dimension is defined by the boundary conditions for the metric, more specifically by the value of its component in the diagonal of the field-free metric. In my notation, a spacelike dimension has positive signature and a timelike dimension has negative signature. (Assume the values in the diagonal are +1 or −1 for convenience: +2 results in the same physics as +1, etc.) It is a remarkable thing that this is the *only* difference between space and time: coordinate invariance ensures that space and time cannot otherwise be distinguished in any law of physics. With a degenerate metric there can be a zero in the diagonal of the field-free metric. This zero corresponds to the so-called lightlike dimension and is distinct from space and time because it is not possible to transform to it. (Indeed, this is the only other distinct possibility for a type of dimension, ignoring possible complex dimensions like complex time.) I will argue later that such a dimension will not be directly observable, but in curved space it will result in a force which may be interpreted as electromagnetism.

There are three possible degenerate theories: one where the covariant metric is degenerate, one where the contravariant metric is degenerate, and one where both metrics are degenerate. The latter is the one that is of primary interest, however all three will be considered.

Let indices $a$, $b$, $c$, ... run over 0, 1, 2, 3, 5 and let $\mu$, $\nu$, $\rho$, ... run over 0, 1, 2, 3 and denote the fifth dimension by $w$. Denote the covariant metric by $\gamma_{ab}$ (replacing $g$ with $\gamma$ in five dimensions) and the contravariant metric by $\gamma^{ab}$. If the covariant metric is degenerate then it obeys

$$\det \gamma_{ab} = 0. \tag{1}$$

This will hold in all frames if it holds in one frame even though the left-hand side is not a scalar. In the absence of forces the metric is $\eta_{ab}$ = diag (−1, 1, 1, 1, 0). Because equation (1) is required to hold everywhere it is raised to the status of a law of physics. Now that the determinant of the covariant metric is zero, it is no longer possible to define a contravariant (inverse) metric by

$$\gamma^{ab}\gamma_{bc} = \delta^a_c. \tag{2}$$

Instead the contravariant metric can be partially defined by letting it satisfy

$$\gamma_{ab}\gamma^{bc}\gamma_{cd} = \gamma_{ad}. \tag{3}$$

If $\det \gamma_{ab} \neq 0$ then equation (2) (and general relativity) is recovered trivially. However equation (3) is also consistent with $\det \gamma_{ab} = 0$ because both sides of the equation then have determinant zero. The contravariant metric is not completely defined, for let $\varepsilon^b$ satisfy

$$\gamma_{ab}\varepsilon^b = 0.$$

Then if $\gamma^{ab}$ satisfies (3), so does $\gamma^{ab} + (\alpha^a\varepsilon^b + \alpha^b\varepsilon^a)$ for any vector $\alpha^a$, so the contravariant metric has five as yet undefined components. The existence of an eigenvector with eigenvalue zero is implied by equation (1): note that $\varepsilon^b$ is only defined up to an overall scalar function. One of the five remaining components can be defined by making the contravariant metric degenerate also. It then satisfies

$$\gamma^{ab}\gamma_{bc}\gamma^{cd} = \gamma^{ad}. \qquad (4)$$

The other four components will be defined by a differential equation rather than algebra. I call equations (3) and (4) together the *metric equations*. Let $\varepsilon_b$ be the eigenvector with lowered index:

$$\gamma^{ab}\varepsilon_b = 0.$$

Now that the covariant and contravariant metrics are not inverses of each other it becomes necessary to think of them as two separate but related objects, rather than as a single object with either raised or lowered indices according to preference. Indices must be raised and lowered explicitly in this theory, for consider a contravariant vector $U^a$. If one defined a corresponding covariant vector $U_a = \gamma_{ab}U^b$, then raising the index again would give $\gamma^{ab}U_b = \gamma^{ab}\gamma_{bc}U^c$ which may be different from the original vector and therefore cannot be written as $U^a$. Define

$$h_b^a = \gamma^{ac}\gamma_{cb},$$

a tensor. $h_b^a$ is not equal to $\delta_b^a$. Note that no equation may explicitly refer to the fifth dimension, so there is no condition $\partial_5\gamma_{ab} = 0$: all fields are potentially allowed to vary with $w$ (unlike Kaluza-Klein theory).

If the covariant metric is degenerate but the contravariant metric is not, then the alternative is to require the laws of physics to be invariant under the transformation $\gamma^{ab} \rightarrow \gamma^{ab} + \varepsilon^a\varepsilon^b$, which I call a *metric transformation*. This has the effect of rendering one component of the contravariant metric irrelevant to physics. Likewise if the contravariant metric is degenerate but the covariant metric is not, then one uses the invariance $\gamma_{ab} \rightarrow \gamma_{ab} + \varepsilon_a \varepsilon_b$. As I have already stated, I will primarily consider the case where both metrics are degenerate, however I will argue in the discussion of the equations of motion that for the purposes of describing electromagnetism the theory with the latter metric transformation is an equivalent approach.

## 3  Tensor Calculus

This section follows the standard approach to the development of tensor calculus for general relativity, but is expanded to include degenerate metrics. Because of the extra layer of complexity the algebra is correspondingly more difficult, even if the techniques are not conceptually too different. Note that if $h_b^a = \delta_b^a$ all the usual quantities that are derived reduce to their counterparts in general relativity.

Begin by defining covariant differentiation. For a contravariant vector $U^b$ define covariant derivative

$$\nabla_a U^b = h_a^c \partial_c U^b + \Gamma_{ac}^b U^c,$$

and for a covariant vector $V_b$ define covariant derivative

$$\nabla_a V_b = h_a^c \partial_c V_b - \Gamma_{ab}^c V_c,$$

and let the product rule hold. Under a coordinate transformation the new Christoffel symbols become

$$\Gamma_{b'c'}^{a'} = P_a^{a'} P_{b'}^b P_{c'}^c \Gamma_{bc}^a - P_{b'}^b P_{c'}^c (\partial_c P_a^{a'}) h_b^a,$$

where $P_a^{a'} = \dfrac{\partial x^{a'}}{\partial x^a}$ is the matrix of derivatives for a coordinate transformation. $\Gamma_{bc}^a$ is not symmetric under $b \leftrightarrow c$. By setting $\nabla_a \gamma_{bc} = 0$ and $\nabla_a \gamma^{bc} = 0$ one finds

$$\Gamma_{bc}^a = \tfrac{1}{2} \gamma^{ad} (h_b^e \partial_e \gamma_{cd} + h_c^e \partial_e \gamma_{bd} + \gamma_{be} \gamma_{cf} \partial_d \gamma^{ef}) + \tfrac{1}{2} h_d^a (\partial_b h_c^d + \partial_c h_b^d)$$
$$- h_b^d h_c^e (\partial_d h_e^a - \partial_e h_d^a) - \partial_c h_b^a.$$

However, it turns out that for some purposes the Christoffel symbols

$$\Gamma^a_{bc} = \tfrac{1}{2}\gamma^{ad}(\partial_b\gamma_{cd} + \partial_c\gamma_{bd} - \partial_d\gamma_{bc}) - \partial_c h^a_b$$

are more convenient to use, and these are the ones I shall adopt. Note that with this choice of symbols neither of the metrics has covariant derivative zero. Covariant differentiation therefore has no physical interpretation, it is merely a method of obtaining a derivative that is a tensor. Define

$$H^a_{bc} = h^a_d \partial_b h^d_c - h^a_d \partial_c h^d_b - h^d_b \partial_d h^a_c + h^d_c \partial_d h^a_b,$$

a tensor (antisymmetric under b↔c), and Riemann tensor

$$R^a_{bcd} = h^f_d h^e_c \partial_e \Gamma^a_{bf} - h^f_d h^e_b \partial_e \Gamma^a_{cf} + h^f_d \Gamma^e_{bf}\Gamma^a_{ce} - h^f_d \Gamma^e_{cf}\Gamma^a_{be} + h^f_d \Gamma^a_{ef}(\partial_b h^e_c - \partial_c h^e_b) - H^e_{bc}\Gamma^a_{de}$$

(antisymmetric under b↔c). $R^a_{bcd}$ is a tensor since

$$h^e_d \nabla_b \nabla_c V_e - h^e_d \nabla_c \nabla_b V_e = -H^a_{bc}(h^e_d \partial_a V_e - \Gamma^e_{da}V_e) + R^e_{bcd}V_e$$

and $h^e_d \partial_a V_e - \Gamma^e_{da}V_e$ is a tensor. As usual define Ricci tensor

$$R^a_c = R^a_{bcd}\gamma^{bd}$$

and Ricci scalar

$$R = R^a_a.$$

Because both metrics are degenerate a new field $\varphi$ has to be introduced to replace $\sqrt{g}$ so that densities can be integrated. Under a coordinate transformation $\varphi$ becomes $\varphi' = (\det P^a_{a'})\varphi$ and so $\varphi d^5 x$ is a scalar. Define

$$\Phi_a = h^c_a \frac{(\partial_c \varphi)}{\varphi} - \Gamma^c_{ac},$$

a vector, and $\Phi = \Phi_a \Phi_b \gamma^{ab}$. The $\varphi$ field plays a similar role to the scalar field in Brans-Dicke theory [6], except, of course, it is not a scalar.

The question now arises as to how many scalars there are in total, so that the most general Lagrangian can be constructed. To answer this it is necessary to find a set of linearly independent scalars which forms a basis for all scalars. A simple (if brutal) method is to construct the scalars directly from the metrics (instead of using some of the derived quantities, e.g. the Christoffel symbols). Assuming that we are only interested in scalars with two derivatives, each one is then a linear combination of the following possible terms:

$$(\partial_a \gamma_{bc})(\partial_d \gamma_{ef}) \quad (\partial_a \gamma_{bc})(\partial_d \gamma^{ef}) \quad (\partial_a \gamma^{bc})(\partial_d \gamma^{ef}) \quad \partial_a \partial_b \gamma_{cd} \quad \partial_a \partial_b \gamma^{cd}$$

$$\frac{(\partial_a \varphi)}{\varphi}\frac{(\partial_b \varphi)}{\varphi} \quad \frac{(\partial_a \varphi)}{\varphi}(\partial_b \gamma_{cd}) \quad \frac{(\partial_a \varphi)}{\varphi}(\partial_b \gamma^{cd}) \quad \frac{\partial_a \partial_b \varphi}{\varphi}$$

with all possible contractions using $\gamma^{ab}$, $\delta^a_b$, $h^a_b$ and $\gamma_{ab}$. In all there are forty-nine possible terms. However, using the metric equations twenty-six can be eliminated as being linearly dependent on the twenty-three that remain. For example, it is easy to prove that

$$\gamma^{ab}(\partial_c \gamma_{ab}) = -\gamma_{ab}(\partial_c \gamma^{ab}),$$

i.e. $\quad \partial_c(\gamma^{ab}\gamma_{ab}) = 0$.

By transforming each of the twenty-three terms and solving the resulting simultaneous equations three scalars are found:

$$(\partial_a \partial_b \gamma_{cd})\gamma^{ac}\gamma^{bd} - (\partial_a \partial_b \gamma_{cd})\gamma^{ab}\gamma^{cd} + (\partial_a \gamma_{bc})(\partial_d \gamma^{ed})h^a_e \gamma^{bc} - (\partial_a \gamma_{bc})(\partial_d \gamma^{ad})\gamma^{bc}$$
$$- (\partial_a \gamma_{bc})(\partial_d \gamma_{ef})\gamma^{ab}\gamma^{de}\gamma^{cf} + 2(\partial_a \gamma_{bc})(\partial_d \gamma^{cd})\gamma^{ab} - 2(\partial_a \gamma_{bc})(\partial_d \gamma^{ed})\gamma^{ab}h^c_e$$
$$- 2(\partial_a \gamma_{bc})(\partial_d \gamma^{eb})h^a_e \gamma^{cd} + (\partial_a \gamma_{bc})(\partial_d \gamma^{ab})\gamma^{cd} + (\partial_a \gamma_{bc})(\partial_d \gamma_{ef})\gamma^{ab}\gamma^{cd}\gamma^{ef}$$
$$- \tfrac{1}{2}(\partial_a \gamma_{bc})(\partial_d \gamma^{bc})\gamma^{ad} - \tfrac{3}{2}(\partial_a \gamma_{bc})(\partial_d \gamma_{ef})\gamma^{ae}\gamma^{bd}\gamma^{cf}$$
$$+ \tfrac{1}{4}(\partial_a \gamma^{bc})(\partial_d \gamma^{ef})\gamma^{ad}\gamma_{be}\gamma_{cf} - \tfrac{1}{4}(\partial_a \gamma_{bc})(\partial_d \gamma_{ef})\gamma^{ad}\gamma^{bc}\gamma^{ef}$$

$$\gamma^{ab}\frac{(\partial_a \varphi)}{\varphi}\frac{(\partial_b \varphi)}{\varphi} + 2\frac{(\partial_a \varphi)}{\varphi}(\partial_b \gamma_{cd})\gamma^{ac}\gamma^{bd} - \frac{(\partial_a \varphi)}{\varphi}(\partial_b \gamma_{cd})\gamma^{ab}\gamma^{cd}$$
$$+ 2\frac{(\partial_a \varphi)}{\varphi}(\partial_b \gamma^{cb})h^a_c - (\partial_a \gamma_{bc})(\partial_d \gamma^{ed})h^a_e \gamma^{bc} - (\partial_a \gamma_{bc})(\partial_d \gamma_{ef})\gamma^{ab}\gamma^{cd}\gamma^{ef}$$
$$+ (\partial_a \gamma_{bc})(\partial_d \gamma_{ef})\gamma^{ab}\gamma^{de}\gamma^{cf} + 2(\partial_a \gamma_{bc})(\partial_d \gamma^{ed})\gamma^{ab}h^c_e + (\partial_a \gamma^{ac})(\partial_d \gamma^{df})\gamma_{cf}$$
$$+ \tfrac{1}{4}(\partial_a \gamma_{bc})(\partial_d \gamma_{ef})\gamma^{ad}\gamma^{bc}\gamma^{ef}$$

$$\gamma^{ab}\frac{(\partial_a\partial_b\varphi)}{\varphi} + \frac{(\partial_a\varphi)}{\varphi}(\partial_b\gamma^{ab}) + \frac{(\partial_a\varphi)}{\varphi}(\partial_b\gamma^{cb})h_c^a + \frac{(\partial_a\varphi)}{\varphi}(\partial_b\gamma_{cd})\gamma^{ac}\gamma^{bd}$$

$$-\tfrac{1}{2}\frac{(\partial_a\varphi)}{\varphi}(\partial_b\gamma_{cd})\gamma^{ab}\gamma^{cd} + \tfrac{1}{2}(\partial_a\partial_b\gamma_{cd})\gamma^{ac}\gamma^{bd} - \tfrac{1}{2}(\partial_a\partial_b\gamma_{cd})\gamma^{ab}\gamma^{cd} + \tfrac{1}{2}(\partial_a\partial_b\gamma^{ab})$$

$$+ \tfrac{1}{2}(\partial_a\partial_b\gamma^{ab}) - \tfrac{1}{2}(\partial_a\gamma_{bc})(\partial_d\gamma^{bc})\gamma^{ad} - \tfrac{1}{2}(\partial_a\gamma_{bc})(\partial_d\gamma^{ad})\gamma^{bc}$$

$$+ (\partial_a\gamma_{bc})(\partial_d\gamma^{cd})\gamma^{ab} + \tfrac{1}{2}(\partial_a\gamma^{ac})(\partial_d\gamma^{df})\gamma_{cf} - \tfrac{1}{2}(\partial_a\gamma^{dc})(\partial_d\gamma^{af})\gamma_{cf}.$$

The first scalar is the same as the Ricci scalar derived above, the second scalar is $\Phi$ and the third is the divergence $\varphi^{-1}\partial_a(\varphi\gamma^{ab}\Phi_b)$ which can be ignored as a term in the Lagrangian. The most general Lagrangian can therefore be written as

$$L = \frac{c^3}{16\pi G}(R - \omega\Phi - 2\Lambda) \tag{5}$$

where $\Lambda$ is the cosmological constant and $\omega$ is constrained to be greater than the order of 500 by observation [7]. This theory cannot strictly be considered to be a unified theory if one demands that there be a single Lagrangian, however only the $\varphi$ field stands apart. A unique Lagrangian can be obtained by requiring invariance under the transformation $\gamma_{ab} \to \lambda\gamma_{ab}$, $\gamma^{ab} \to \lambda^{-1}\gamma^{ab}$, $\varphi \to \lambda\varphi$ (conformal invariance), when $\omega = -\tfrac{3}{2}$ and $\Lambda = 0$, however this is only consistent with massless fields.

In the theory with degenerate contravariant metric and non-degenerate covariant metric there is a fourth scalar $H = H_{bc}^a H_{ef}^d \gamma_{ad}\gamma^{be}\gamma^{cf}$ which would be combined with the Ricci scalar in a Lagrangian since only the combination $R + \tfrac{3}{2}H$ is invariant under the metric transformation $\gamma_{ab} \to \gamma_{ab} + \varepsilon_a\varepsilon_b$.

## 4  The Equations of Motion

The first approach to finding equations of motion is to use Hamilton's method. Consider a particle moving along a path $x^a(s)$. The equations of motion are obtained in the usual way from Hamilton's equations $\frac{dx^a}{ds} = \frac{\partial H}{\partial p_a}$ and $\frac{dp_a}{ds} = -\frac{\partial H}{\partial x^a}$ with Hamiltonian $H = \frac{1}{2mc}\gamma^{ab}p_a p_b$. One finds

$$mcu^a = \gamma^{ab}p_b \tag{6}$$

and
$$\frac{dp_a}{ds} = -\frac{1}{2mc}(\partial_a \gamma^{bc}) p_b p_c \tag{7}$$

where $u^a = \frac{dx^a}{ds}$. It follows from (6) and (7) that $u^a p_a$ = constant and $u^a \nabla_a p_b = 0$. These equations can be related back to four-dimensional relativity and Maxwell's theory by using a notation where the $w$-coordinate is split out from the four space-time coordinates. Write

$$\gamma^{ab} = \begin{pmatrix} g^{\mu\nu} & -\kappa g^{\mu\rho} A_\rho \\ -\kappa g^{\nu\sigma} A_\sigma & \kappa^2 g^{\rho\sigma} A_\rho A_\sigma \end{pmatrix} \tag{8}$$

where $\kappa$ is a constant, and let $F_{\mu\nu} = \partial_\mu A_\nu - \partial_\nu A_\mu$. Then if $g_{\mu\nu} = \eta_{\mu\nu}$ and $\partial_5 \gamma^{ab} = 0$ equations (6) and (7) reduce to $p_5$ = constant and

$$mc\frac{du^\mu}{ds} = \kappa F^\mu{}_\nu u^\nu p_5$$

where indices have been raised in the traditional four-dimensional manner, i.e. the familiar form from Maxwell's theory, provided that $p_5 = \kappa^{-1} q$. Identify $A_\mu$ with electromagnetism. The coordinate transformation $w \to w + f(x^\mu)$ is equivalent to a gauge transformation, under which $A_\mu \to A_\mu - \kappa^{-1} \partial_\mu f$. Note that $\varepsilon_\mu = \kappa A_\mu \varepsilon_5$, so the electromagnetic potential is identifiable in five dimensions up to an overall scalar function.

Equation (6) implies that $\varepsilon_a u^a = 0$, so that (locally) there is a direction in which the coordinate does not change. I postulate that this is the reason why we do not notice the extra dimension directly. The null direction can change from point to point resulting in a force (electromagnetism). The fifth component of momentum can be non-trivial and is identified with charge.

The second approach to finding equations of motion, using the Lagrangian formulation, leads to a different result from Hamilton's method. This does not constitute an error: there are simply two possible equations of motion. With Lagrangian $L = \tfrac{1}{2} mc \gamma_{ab} u^a u^b$ the Lagrange equations

$$\frac{d}{ds}\frac{\partial L}{\partial u^a} - \frac{\partial L}{\partial x^a} = 0$$

give

$$\gamma_{ab}\frac{du^b}{ds}+(\partial_c\gamma_{ab})u^b u^c-\tfrac{1}{2}(\partial_a\gamma_{bc})u^b u^c=0. \tag{9}$$

Let $p_a=mc\gamma_{ab}u^b$: as before $u^a p_a$ = constant. Equation (9) can be written in more familiar form as

$$\frac{d}{ds}(h_b^a u^b)+\Gamma^a_{bc}u^b u^c=0$$

(four equations) plus the contraction of (9) with $\varepsilon^a$:

$$\varepsilon^a u^b u^c(\partial_c\gamma_{ab}-\tfrac{1}{2}\partial_a\gamma_{bc})=0. \tag{10}$$

Continuing the (4+1)-dimensional notation write

$$\gamma_{ab}=\begin{pmatrix} g_{\mu\nu}+\kappa^2\beta(B_\mu A_\nu+B_\nu A_\mu) & \kappa\beta(B_\mu+A_\mu) \\ \kappa\beta(B_\nu+A_\nu) & 2\beta \end{pmatrix}$$

with $\beta$ arranged so that $\det\gamma_{ab}=0$, i.e.

$$\beta=\frac{2}{\kappa^2 g^{\mu\nu}(B_\mu-A_\mu)(B_\nu-A_\nu)}.$$

If $g_{\mu\nu}=\eta_{\mu\nu}$, $A_\mu=0$ and $\partial_5\gamma_{ab}=0$ one finds

$$\frac{dp_\mu}{ds}=\kappa\partial_\mu(\beta B_\nu)g^{\nu\rho}p_\rho u^5$$

and $\kappa\beta B_\mu u^\mu+2\beta u^5$ = constant.

From the above there are three reasons why it looks wrong to use the Lagrange equations. First, equation (10) imposes a restriction on the boundary conditions which looks difficult to satisfy generally. Second, there is no equation like $\varepsilon_a u^a=0$ to explain why the fifth dimension is not observable (only $\varepsilon^a p_a=0$). And third, the equations of motion cannot be related back to Maxwell's theory, making a physical interpretation difficult. It is clear therefore that the most interesting (classical) equations of motion are the ones derived from a Hamiltonian.

This settles the case for the contravariant metric being degenerate, but not the covariant metric. However, if $\det \gamma_{ab} \neq 0$ is allowed it will only be in the context of a theory which is invariant under the metric transformation, so given non-degenerate $\gamma_{ab}$ it will always be possible to find some equivalent $\tilde{\gamma}_{ab} = \gamma_{ab} + \varepsilon_a \varepsilon_b$ such that $\det \tilde{\gamma}_{ab} = 0$, which is the same as having a degenerate covariant metric by principle. I work with both metrics degenerate for definiteness.

Although the degenerate theory describes the physics of a dimension which could be said to have zero size (a lightlike dimension), and Kaluza-Klein theory describes a "curled-up" dimension which has a small but definite non-zero size, it is not possible to think of the lightlike dimension as the limit of a curled-up dimension as its size tends to zero. For consider the Kaluza-Klein contravariant metric

$$\gamma^{ab} = \begin{pmatrix} g^{\mu\nu} & -\kappa g^{\mu\rho} A_\rho \\ -\kappa g^{\nu\sigma} A_\sigma & \kappa^2 g^{\rho\sigma} A_\rho A_\sigma + \lambda \end{pmatrix}$$

where $\lambda$ is some parameter which roughly corresponds to the size of the extra dimension. This metric tends to (8) in the limit as $\lambda$ tends to zero. However the corresponding covariant metric

$$\gamma_{ab} = \begin{pmatrix} g_{\mu\nu} + \lambda^{-1} \kappa^2 A_\mu A_\nu & \lambda^{-1} \kappa A_\mu \\ \lambda^{-1} \kappa A_\nu & \lambda^{-1} \end{pmatrix}$$

diverges in the limit. Indeed, it should be obvious that whereas Kaluza-Klein theory describes one spin-1 field the degenerate theory describes two, so one cannot be the limit of the other.

## 5  The Field Equations

The next step in the construction of the theory is to find the field equations. There will be nineteen equations in all: in terms of the (4+1)-dimensional notation of the previous section ten equations for $g_{\mu\nu}$, four each for $A_\mu$ and $\beta B_\mu$, and one for $\varphi$. Begin with the vacuum field equations, obtained by varying the Lagrangian (5):

$$\delta \int L\varphi \, d^5 x = 0 .$$

The equation for $A_\mu$ can be written in terms of the tensor $E_{ab} = \partial_a \varepsilon_b - \partial_b \varepsilon_a$ as

$$(\nabla_a E_{bc})\gamma^{ab}\gamma^{ce} + 2H^a_{bc}E_{ad}\gamma^{bd}\gamma^{ce} + (2\omega+3)\Phi_a E_{bc}\gamma^{ab}\gamma^{ce} = 0. \tag{11}$$

This represents four equations, since contracting with $\varepsilon^e$ gives a trivial equation. The terms in (11) are arranged so that it is invariant under $\varepsilon_b \to \lambda\varepsilon_b$. If $g_{\mu\nu} = \eta_{\mu\nu}$, $\varphi = 1$ and $\partial_5$ fields $= 0$ then this becomes

$$\partial_\mu F^\mu{}_\sigma - \tfrac{1}{2}\kappa^2\beta(B_\sigma - A_\sigma)F_{\mu\nu}F^{\mu\nu} + (2\omega+1)\kappa^2\beta(B_\mu - A_\mu)F^{\mu\rho}F_{\rho\sigma} = 0.$$

Ignoring the non-linear terms this equation has the same form as Maxwell's theory.

The next equation to write is also a vector equation: the equation for $\varepsilon^c$. Define symmetric field tensor $E^{ab} = \gamma^{ac}(\partial_c\varepsilon^b) + \gamma^{bc}(\partial_c\varepsilon^a) - \varepsilon^c(\partial_c\gamma^{ab})$ and corresponding scalar $E = \gamma_{ab}E^{ab}$. Then

$$(\nabla_a E^{ac})h^e_c - (\partial_a E)\gamma^{ae} + 2(\nabla_a\gamma_{bc})E^{ab}\gamma^{ce} - 2(\nabla_a h^a_b)E^{bc}h^e_c$$
$$- \Phi_a h^a_b E^{bc}h^e_c + \Phi_a E\gamma^{ae} + 2\omega\varepsilon^a(\partial_a\Phi_c - \Gamma^b_{ca}\Phi_b)\gamma^{ce} = 0. \tag{12}$$

As above this equation is invariant under $\varepsilon^c \to \lambda\varepsilon^c$. If $g_{\mu\nu} = \eta_{\mu\nu}$, $\varphi = 1$, $A_\mu = 0$ and $\partial_5$ fields $= 0$ then (12) becomes

$$\partial^\mu(\partial_\mu\beta B_\nu - \partial_\nu\beta B_\mu) = 0.$$

This too has the form of Maxwell's theory, however the field cannot be interpreted as electromagnetism because the equations of motion come out wrongly. Also, it is difficult to imagine a coordinate transformation which would correspond to a gauge transformation for $\beta B_\mu$.

The field equations for gravitation are

$$h^a_c\gamma^{bd}R^c_d + h^a_c E^{cd}h^e_d E_{ef}\gamma^{bf} - (\nabla_c\Phi_d)\gamma^{ac}\gamma^{bd} - (\omega+1)\Phi_c\Phi_d\gamma^{ac}\gamma^{bd}$$
$$+ (\omega+1)\frac{1}{\varphi}\partial_c(\varphi\gamma^{cd}\Phi_d)\gamma^{ab} + a\leftrightarrow b = 0 \tag{13}$$

where the convention is that where $\varepsilon_a$ and $\varepsilon^a$ appear in the same equation they are normalized so that $\varepsilon_a\varepsilon^a = 1$. There are ten equations: twenty-five for the two indices (five times five) minus ten for the symmetry under $a\leftrightarrow b$, minus five trivial equations from contracting with $\varepsilon_a$ or $\varepsilon_b$.

The final field equation, for $\varphi$, is

$$\frac{2\omega}{\varphi}\partial_c(\varphi\gamma^{cd}\Phi_d) + R - \omega\Phi - 2\Lambda = 0. \qquad (14)$$

Now introduce matter as a source for the field. Since the Lagrange equations of motion have been demonstrated to be unphysical, assume that we are interested in a particle which is described by a Hamiltonian, i.e. equations (6) and (7). The source for gravitation is straightforward, the right-hand side of equation (13) becoming

$$\frac{16\pi G}{c^3}mc\rho u^a u^b$$

where $\rho$ is the particle density. The complications arise with the spin-1 fields. The particle will act as a source for $\beta B_\mu$, the right-hand side of (12) becoming

$$\frac{16\pi G}{c^3}\rho(\varepsilon^c p_c)u^b.$$

However, because $\varepsilon_b u^b = 0$ it is not possible to find a right-hand side for equation (11), so there is no source for electromagnetism. In contrast, the particle moves in the electromagnetic field, but does not "feel" $\beta B_\mu$. (The opposite would be true for a particle described by a Lagrangian. The source for electromagnetism would be proportional to $\rho(\varepsilon_b u^b)p_c$ but there would be no right-hand side for equation (12). The particle would feel $\beta B_\mu$ but not electromagnetism.)

As there is no source for electromagnetism equation (11) has a trivial solution: $E_{ab} = 0$. This poses a problem since it is clearly at odds with the observation that charges interact with each other. However, because there are non-linear terms in (11) there is a new possibility: that electromagnetism can act as its own source, which I will investigate in the next section. Notwithstanding this issue (Coulomb's law is yet to be retrieved), the theory is now a complete theory with equations of motion and field equations.

For the particle define energy-momentum-charge tensor $T_b^a = mc\rho u^a p_b$. Then using the continuity equation $\varphi^{-1}\partial_a(\varphi\rho u^a) = 0$ and the equations of motion $u^a\nabla_a p_b = 0$ the five conservation laws are $(\nabla_a + \Phi_a)T_b^a = 0$.

# 6 The Spherically Symmetric Case

Let us now attempt to find some spherically symmetric solutions to the vacuum field equations. This will provide an opportunity to confirm one of the results from general relativity by retrieving the Schwarzschild solution, and is a simple way to investigate the non-linearity of the spin-1 fields and Coulomb's law: the last step in the theory.

Write the covariant metric as

$$\begin{pmatrix} -B(1-\psi\chi)^2 & 0 & 0 & 0 & \chi B(1-\psi\chi) \\ 0 & A & 0 & 0 & 0 \\ 0 & 0 & r^2 & 0 & 0 \\ 0 & 0 & 0 & r^2 \sin^2\theta & 0 \\ \chi B(1-\psi\chi) & 0 & 0 & 0 & -\chi^2 B \end{pmatrix}$$

and the contravariant metric as

$$\begin{pmatrix} -B^{-1} & 0 & 0 & 0 & \psi B^{-1} \\ 0 & A^{-1} & 0 & 0 & 0 \\ 0 & 0 & r^{-2} & 0 & 0 \\ 0 & 0 & 0 & r^{-2} \sin^{-2}\theta & 0 \\ \psi B^{-1} & 0 & 0 & 0 & -\psi^2 B^{-1} \end{pmatrix}$$

using spherical polar coordinates $(ct, r, \theta, \phi, w)$. Assume that the solution is static and independent of the fifth coordinate and let the cosmological constant be zero. Then

$$R = -\frac{B''}{AB} + \frac{(B')^2}{2AB^2} + \frac{A'B'}{2A^2B} + \frac{2}{r^2}\left[1 - \frac{1}{A} + \frac{r}{A}\left(\frac{A'}{A} - \frac{B'}{B}\right)\right]$$
$$+ 2\frac{\psi''\chi}{A} + 3\frac{\psi'\chi'}{A} - \frac{(\psi')^2\chi^2}{A} - \psi'\chi\frac{A'}{A^2} + 2\psi'\chi\frac{B'}{AB} + 4\frac{\psi'\chi}{Ar}$$

and

$$\Phi = \frac{1}{A}\left(\frac{\varphi'}{\varphi} - \frac{2}{r} - \frac{A'}{2A} - \frac{B'}{2B} + \psi'\chi\right)^2$$

where $\varphi$ has been replaced with $\varphi \sin\theta$ for convenience. For simplicity consider the limit $\omega \to \infty$ in which general relativity is retrieved. Then $\Phi = 0$ and so

$$\frac{\varphi'}{\varphi} = \frac{2}{r} + \frac{A'}{2A} + \frac{B'}{2B} - \psi'\chi. \tag{15}$$

Using (15) equations (11), (12), (13) and (14) imply

$$R = 0 \tag{16}$$

$$R_\theta^\theta = R_\phi^\phi = \frac{1}{r^2}\left[1 - \frac{1}{A} + \frac{r}{A}\left(\frac{A'}{2A} - \frac{B'}{2B} + \psi'\chi\right)\right] = 0 \tag{17}$$

$$-\frac{B''}{2B} + \frac{(B')^2}{4B^2} + \frac{A'B'}{4AB} + \frac{A'}{Ar} + \psi''\chi + 2\psi'\chi' - \psi'\chi\frac{A'}{2A} + \psi'\chi\frac{B'}{B} = 0 \tag{18}$$

$$\psi'' - \psi'\left(\frac{A'}{2A} + \frac{B'}{2B} - \frac{2}{r} + \psi'\chi\right) = 0 \tag{19}$$

and

$$(\chi B)'' - (\chi B)'\left(\frac{A'}{2A} + \frac{B'}{2B} - \frac{2}{r} + \psi'\chi\right) = 0. \tag{20}$$

Equations (19) and (20) can both be integrated once to give

$$\psi' = \frac{AB}{\varphi} \times \text{constant} \tag{21}$$

and

$$(\chi B)' = \frac{AB}{\varphi} \times \text{constant}. \tag{22}$$

The Schwarzschild solution can be retrieved easily by setting $\psi' = 0$. Then $AB = 1$, $\varphi = r^2$ and

$$A = \left(1 - \frac{2Gm}{c^2 r}\right)^{-1}.$$

If $\psi' \neq 0$ consider the case where there is no mass, i.e. $m = 0$. A further integration can be done by combining (21) and (22) and using the boundary conditions $\psi \to 0$ and $\chi \to 0$ as $r \to \infty$ to obtain

$$\chi B = -b^2 \psi \tag{23}$$

where $b$ is a positive constant. Equations (16), (17) and (18) can be combined to give

$$\psi'\chi'+(\psi')^2\chi^2+\frac{1}{r}\left(\frac{A'}{A}+\frac{B'}{B}-2\psi'\chi\right)=0. \tag{24}$$

The five equations to solve are now (15), (17), (21), (23) and (24). From this point there is no systematic way to proceed, however it turns out that more progress can be made if

$$\frac{B'}{2B}=\psi'\chi. \tag{25}$$

Three more integrations can then be done easily. One finds by substituting (23) into (25) that

$$B=1-b^2\psi^2,$$

by substituting (25) into (15) that

$$\varphi=r^2 A^{1/2},$$

and by substituting (25) into (17) that

$$A=\left(1+\frac{a^2}{r^2}\right)^{-1} \tag{26}$$

where $a$ is a positive constant. Equation (24) then becomes

$$\frac{b\psi'}{1-b^2\psi^2}=\pm\sqrt{2}\,\frac{a}{r^2}\left(1+\frac{a^2}{r^2}\right)^{-1/2}. \tag{27}$$

This is the same as equation (21) with the constant equal to $\pm\sqrt{2}a/b$. Since these equations are consistent with each other the guess (25) was valid and we can proceed. Note that equation (27) has been obtained by taking a square root: given (26) the right-hand side of (23) must have a minus sign; if $A$ were $(1-a^2/r^2)^{-1}$ then (23) would have to be $\chi B=+b^2\psi$. Equation (27) can be integrated using partial fractions on the left-hand side and the substitution $\sinh\alpha=a/r$ on the right-hand side. One finds

$$\frac{1+b\psi}{1-b\psi}=\exp\left(\pm 2\sqrt{2}\sinh^{-1}\left(\frac{a}{r}\right)\right)$$

which can be simplified to

$$\pm b\psi = \frac{\left(\dfrac{a}{r} + \sqrt{1 + \dfrac{a^2}{r^2}}\right)^{2\sqrt{2}} - 1}{\left(\dfrac{a}{r} + \sqrt{1 + \dfrac{a^2}{r^2}}\right)^{2\sqrt{2}} + 1}.$$

$\psi$ goes like $\pm\sqrt{2}a/br$ where $r$ is much greater than $a$ and so is a Coulomb potential (Figure 1), however $\psi \to \pm b^{-1}$ as $r \to 0$ and $\psi$ is in fact regular everywhere, so there is no infinity of self-action from the electromagnetic field. The plus or minus corresponds to positive and negative charges. The value for $a$ is $\sqrt{G\mu_0 q^2 / 2\pi c^2}$ which is approximately $10^{-36}$ m for an electron: this is the very small length scale at which the degenerate electromagnetic potential differs from the Coulomb potential. (Generally there will be differences under extreme conditions which will include charged black holes.) The value for $b$ is unimportant and it can be set to unity with the transformation $w \to bw$.

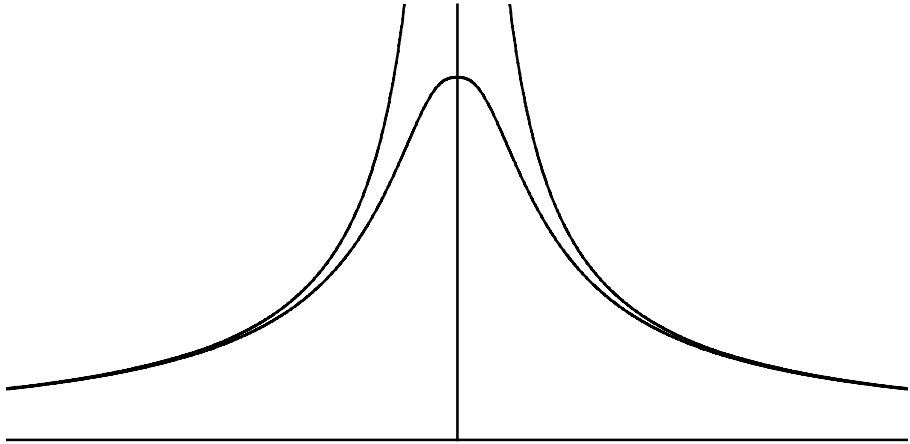

Figure 1: A plot of the degenerate electromagnetic potential (lower line) against the Coulomb potential (upper line).

Like charges repel with a minus on the right-hand side of (23). If it were a plus instead so that like charges attracted, then $A$ would be $(1 - a^2/r^2)^{-1}$ and therefore singular at $r = a$. Assuming that this is a genuine singularity and not (as for the Schwarzschild solution) one which can be removed with a suitable coordinate transformation, then this is a possible argument for why electromagnetism is repulsive.

# 7 Conclusion

It has been shown in this paper that it is possible to develop a degenerate extension to relativity using traditional methods and in a logical and essentially unique way. In five dimensions electromagnetism can be incorporated into the geometry of space-time and unified with gravitation, and as one might expect from a unified theory electromagnetism is in general non-linear. There is a natural explanation as to why the fifth dimension cannot be measured directly, without the need for compactification or any extra conditions. The theory is invariant under all coordinate transformations and general relativity is retrieved in the limit $\omega \to \infty$. The electromagnetic potential is regular everywhere and is close to the Coulomb potential at all except very small length scales.

Despite these positive results, questions will inevitably be asked as to whether the degenerate theory is an improvement on the Kaluza-Klein and Einstein-Maxwell theories. Unification is desirable but not necessary: there are no anomalies like the perihelion advance of Mercury to explain. It is hard to imagine a prediction of this theory being verified experimentally since the differences from Maxwell's theory occur at such small length scales. Small differences are a double-edged sword: they are small enough to mean that the theory is consistent with current experiment, but so small as to be potentially undetectable by any future experiment. Nevertheless my feeling is that the degenerate theory is sufficiently compelling conceptually to warrant further investigation and even acceptance.

# References

[1] Kaluza, T. (1921). *Sitz. Preuss. Akad. Wiss. Phys. Math.*, 966-972
[2] Klein, O. (1926). *Zeitschr. Phys.* **37**, 895-906
[3] Ashtekar, A. (1986). *Phys. Rev. Lett.* **57**, 2244-2247
[4] Bengtsson, I. & Jacobson, T. (1997). *Class. Quantum Grav.* **14**, 3109-3121
[5] Cabral, L.A. & Rivelles, V.O. (2000). *Class. Quantum Grav.* **17**, 1577-1594
[6] Brans, C. & Dicke, R.H. (1961). *Phys. Rev.* **124**, 925
[7] Will, C.M. (1981). Theory and experiment in gravitational physics (Cambridge University Press, Cambridge)